\documentclass{article}
\usepackage{spconf,amsmath,graphicx}
\usepackage{mathtools}
\usepackage{booktabs}
\usepackage{enumitem}
\usepackage{amssymb}
\usepackage{xcolor}

\makeatletter
\renewcommand\footnotesize{\@setfontsize\footnotesize{9}{11}}
\makeatother

\graphicspath{{fig/}}
\title{Ternary Visible Light Communication Using Event-Based Vision Sensors}

\name{Sotaro Kuremoto$^{1}$, Junya Hara$^{1}$, Hiroshi Higashi$^{2}$, Yuichi Tanaka$^{1}$
  \thanks{This work was supported in part by JSPS KAKENHI under Grant 26H02536.}}
\address{$^{1}$The University of Osaka, Osaka, Japan\qquad $^{2}$Kansai University, Osaka, Japan}

\begin{document}
\ninept
\maketitle
\frenchspacing

\begin{abstract}
This paper proposes a ternary visible light communication method using an event-based vision sensor (EVS) and a liquid crystal display (LCD).
The throughput of optical camera communication (OCC) is limited by the frame rates of conventional frame-based cameras, and EVSs are expected to overcome this limitation because their pixels asynchronously trigger events with polarity in response to brightness changes, with a temporal resolution on the order of microseconds.
However, existing event-based OCC relies on binary signaling that uses only the presence or absence of events and leaves the polarity unused.
The proposed method maps the three brightness transitions---increase, decrease, and no change---to \textit{ternary symbols}, increasing the information carried per symbol.
At the transmitter, the LCD displays a marker in which the transmitted data are encoded; at the receiver, the EVS captures the marker and records the brightness changes as events with positive and negative polarities.
To suppress errors from the timing mismatch between display updates at the transmitter and demodulation at the receiver, we also propose a demodulation algorithm based on frame-transition detection.
The experimental results demonstrate that the proposed method achieves throughput equivalent to $336.5$~kbps, $\sim\!150\%$ higher than the binary-based counterpart.
\end{abstract}


\section{Introduction}\label{sec:intro}
Optical camera communication (OCC) is a visible light communication technology in which a transmitter such as an LED or a display sends information to an image sensor~\cite{saha2015survey}.
OCC can use existing displays and imaging devices without dedicated communication equipment~\cite{le2017survey}.
Visible light also has a short wavelength and does not pass through obstacles. Therefore, a physical barrier prevents information leakage~\cite{securityOCC}.
These features make OCC promising for applications such as indoor positioning~\cite{Eroglu_Indoor_Pos_2015} and underwater communication~\cite{zhou2022design}.

The receiver is usually a camera that captures a marker displayed by the transmitter as a sequence of image frames and recovers the data from them.
The throughput is therefore limited by the camera frame rate, which is typically 30 to 60 fps~\cite{liu2020some,saeed2019optical}.

To address this limitation, OCC using an event-based vision sensor (EVS) has been studied~\cite{tofighi2025survey}.
EVSs provide temporal resolution on the order of $10^{-6}$~s and detect increases and decreases in brightness as events with positive and negative polarities, respectively~\cite{EVS}.

The existing event-based OCC method~\cite{Motion} assigns a bit according to the presence or absence of events triggered by a blinking marker.
However, it does not utilize polarities obtained from EVSs.
Using event polarity for modulation increases the information carried per symbol and could further increase throughput.

This paper proposes a \textit{ternary} visible light communication method using an EVS as a receiver and a liquid crystal display (LCD) as a transmitter.
The LCD displays a two-dimensional marker with pixel-level brightness control.
At the transmitter, the three possible brightness transitions of a marker cell---increase, decrease, and \textit{no change}---represent the data, allowing the marker to carry ternary symbols.
At the receiver, the EVS captures the marker and records the brightness changes as events.

To prevent symbol-decision errors due to the characteristics of LCDs, we introduce a demodulation algorithm that uses brightness changes in the marker as a synchronization signal.
We also place the finder patterns (FPs) used in QR codes~\cite{QRstandard} at the four corners of the marker, and locate the reference points from them with subpixel accuracy to suppress errors from inaccurate localization.

We conducted experiments evaluating the throughput and the error rate.
The proposed method achieves a higher throughput and a lower bit error rate (BER) than the existing binary method, with its highest mean throughput equivalent to $336.5$~kbps.
The FPs further reduce the BER.

\section{Binary OCC Using EVS and LCD}\label{sec:evsocc}
We introduce EVSs and the existing EVS-based OCC system~\cite{Motion}.

\subsection{Event-Based Vision Sensor}\label{sec:evs}
EVSs are sensors whose pixels $\mathbf{u}\in\Omega=\{0,\dots,H-1\}\times\{0,\dots,W-1\}$ detect brightness changes independently and asynchronously~\cite{EVS},
where $H$ and $W$ are the numbers of pixels in the vertical and horizontal directions, respectively.
The brightness change detected by an EVS is expressed as the temporal difference $\Delta L(\mathbf{u},t)=L(\mathbf{u},t)-L(\mathbf{u},t-\Delta t)$ in the log brightness $L(\mathbf{u},t)=\ln I(\mathbf{u},t)$,
where $I(\mathbf{u},t)$ is the brightness of pixel $\mathbf{u}$ at time $t$ and $\Delta t$ is the time since the previous event at $\mathbf{u}$.

An event $e_k=(\mathbf{u}_k,t_k,p_k)$ is triggered when
\begin{equation}
 p_k\,\Delta L(\mathbf{u}_k,t_k)=\left|\Delta L(\mathbf{u}_k,t_k)\right|\ge C,
 \label{eq:evs}
\end{equation}
where $C>0$ is the contrast threshold, and $p_k=+1$ and $p_k=-1$ denote an increase and a decrease in brightness, respectively.
An EVS outputs events only for the pixels and times at which the brightness changes.
The temporal resolution of EVSs is typically on the order of $10^{-6}$~s.
EVSs are beneficial for OCC because of this high temporal resolution.

\subsection{Dynamic Marker and Its Decoding}\label{sec:binary}
\begin{figure}[t]
 \centering
   \includegraphics[width=\columnwidth]{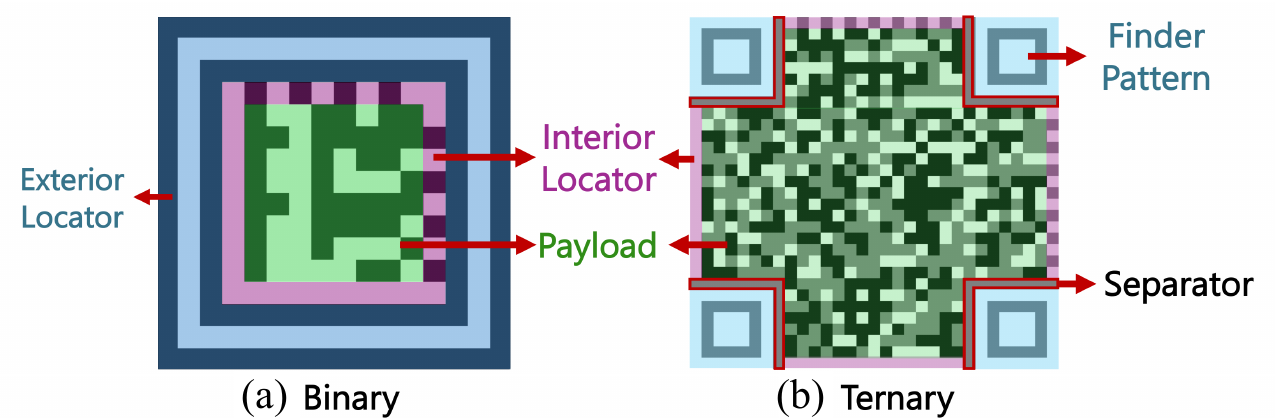} \caption{Structure of (a) the binary dynamic marker of the existing method~\cite{Motion} and (b) the proposed ternary dynamic marker. The payload and the interior locator appear in both markers, whereas the finder patterns and the separators are specific to the proposed marker.}
 \label{fig:tmarker}
 \vspace{-6pt} \end{figure}

In~\cite{Motion}, the transmitter displays a dynamic marker whose brightness changes over time.
As shown in Fig.~\ref{fig:tmarker}(a), the marker consists of a payload for the transmitted data, an interior locator that determines its orientation, and an exterior locator that distinguishes it from the background.

\suppressfloats[t]

The transmitter shows a data frame and a blank frame alternately at a frame rate of $f$ fps.
Each of the $N_d$ payload cells takes the pixel value $255$ or $0$ in a data frame according to the transmitted bit, and the pixel value $0$ in a blank frame.
Data frames are therefore displayed at $f/2$ fps, and the maximum throughput is $fN_d/2$~bps.

\begin{figure}[t]
 \centering
 \includegraphics[width=0.90\columnwidth]{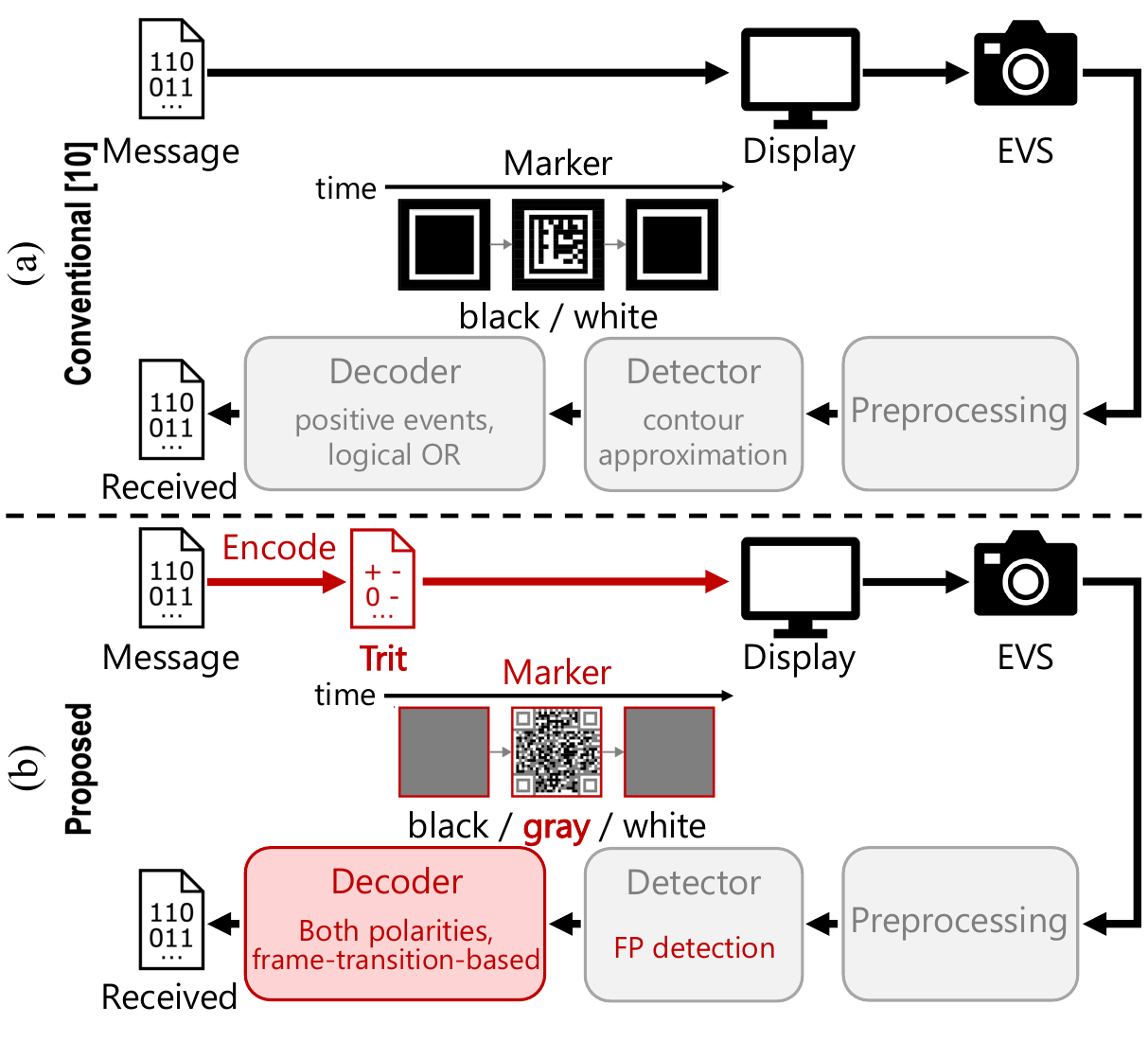} \caption{Encode-decode flows of (a) the existing binary OCC~\cite{Motion} and (b) the proposed ternary OCC. Red parts are newly introduced building blocks that differ from the existing method. The display alternates between a blank frame and a data frame, where gray represents the intermediate brightness.}
 \label{fig:flow}
 \vspace{-6pt} \end{figure}

The receiver captures the dynamic marker displayed on the LCD using the EVS and recovers the transmitted data from the detected events. As shown in Fig.~\ref{fig:flow}(a), the receiver has three stages: preprocessing, detection, and decoding\footnote{We assume a static setting in which the marker and the sensor do not move relative to each other.}\nobreak.

\suppressfloats[t]

\noindent
\textbf{Preprocessing:}\label{sec:preprocessing}
During preprocessing, the receiver applies a refractory filter, which removes repetitive events at the same pixel within a fixed time window, and a spatiotemporal filter, which removes events isolated in space and time~\cite{preprocessing}.

\noindent
\textbf{Marker Detection:}\label{sec:binary_detection}
The detection stage determines the marker from a set of observed events.
In a cell that switches to the pixel value $255$, the transition from a blank frame to a data frame triggers a positive event, and the transition back to a blank frame triggers a negative event.
Events of the same polarity are therefore observed with a period of $2/f$.

The detection stage obtains this periodicity from the surface of active events (SAE)~\cite{SAE}.
Let $T_1$ be the time at which the SAE is evaluated.
The SAE returns the timestamp [s] of the last event of polarity $p$ at each pixel $\mathbf{u}$ before $T_1$:
\begin{equation}
 S_p(\mathbf{u};T_1)=\max\{t_k \mid e_k\in\mathcal{E},\,\mathbf{u}_k=\mathbf{u},\, t_k<T_1,\, p_k=p\},
 \label{eq:sae}
\end{equation}
where $\mathcal{E}=\{e_k\}$ is the set of events.

The detection stage evaluates $S_p$ at successive times and computes the intervals between events of the same polarity at each pixel.
A binary map indicating pixels with event intervals close to $2/f$ is then constructed.
The marker region is identified by contour approximation, which extracts contours from the binary map and approximates them as polygons.

\noindent
\textbf{Decoding:}\label{sec:binary_decoding}
The decoding stage determines the bit in each cell from the positive events observed within that cell.
Because an LCD updates each frame progressively from top to bottom, the rows of the marker change their brightness at different times.
A set of events observed in a short time window may contain no events in some rows.
The decoding stage therefore determines the cell values several times in the data-frame period $2/f$.
It then combines the partial decisions of each cell by a logical OR.

\section{Ternary OCC Using EVS and LCD}\label{sec:proposed}
This section presents the proposed ternary OCC method. Its overview is shown in Fig.~\ref{fig:flow}(b).
The red-colored parts in the figure are differences from~\cite{Motion} while the overall encode-decode flow is similar.
The most important part is that we use \textit{trits} instead of bits to encode data: It is naturally done by using EVS characteristics.
According to the ternary symbol transmission, we propose a specially-designed marker at the transmitter side.
Furthermore, at the receiver side, a new decoder is considered that takes into account ternary symbols along with the robust marker detection.

In the following, we describe the details of the newly-introduced parts for trit-based OCC using EVSs.

\subsection{Data Encoding}\label{sec:encoding}
We introduce the ternary symbols $\{+1,0,-1\}$, which we call trits.
Suppose that we have a bitstream to be transmitted. It is divided into three-bit groups $(b_2,b_1,b_0)$, and each group is converted into two trits $(t_0,t_1)$ as follows:
\begin{equation}
 t_0=(v\bmod 3)-1,\quad t_1=\lfloor v/3\rfloor-1,
 \label{eq:enc}
\end{equation}
where $v=2^2b_2+2^1b_1+2^0b_0$.
Since $v$ ranges from $0$ to $7$, eight out of the nine trit pairs are used, whereas the pair $(+1,+1)$, which corresponds to $v=8$, is unused.
The mapping is one-to-one, and therefore each pair of cells carries three bits: $1.5$ bits per cell.

\subsection{Marker Structure}\label{sec:structure}
Fig.~\ref{fig:tmarker}(b) shows the proposed ternary dynamic marker, a grid of $N_g\times N_g$ cells forming the payload, the finder patterns, the separators, and the interior locator.
Each cell contains $s_c\times s_c$ pixels. The blank frame with the intermediate pixel value is inserted between the data frames to reset the pixel values.
\begin{itemize}[topsep=1pt,itemsep=0pt,parsep=0pt,leftmargin=*]
 \item The \textbf{payload} carries the data: A cell represents $+1$, $0$, or $-1$. From the intermediate pixel value of the blank frame, they are represented by changing to a high brightness level, staying unchanged, or changing to a low brightness level, respectively.
 That is, a high-brightness cell triggers a positive event, a low-brightness cell a negative event, and an unchanged cell no event.
 \item Each \textbf{finder pattern} is a $7\times7$ cell arrangement of concentric squares at one of the four corners, alternating blinking white and static intermediate-brightness cells. Along a scan line through its center, the presence and absence of events form the same $1:1:3:1:1$ ratio as in a QR code~\cite{QRstandard}.
 \item Each \textbf{separator} is a one-cell-wide, L-shaped region on the inner side of an FP. Its brightness does not change between frames, which prevents the payload from interfering with the $1:1:3:1:1$ ratio.
 \item The \textbf{interior locator} is the one-cell-wide region along the marker boundary, excluding the FPs. Two sides are solid white and the other two alternate between the intermediate brightness and white, which uniquely determines the marker orientation.
\end{itemize}

\subsection{Marker Localization by Finder Pattern Detection}\label{sec:detection}
During marker detection, we estimate the four FP centers, which serve as the reference points, using QR-code FP detection~\cite{QR_Belussi_2011,QR_Szentandrasi_2012}.
We scan the binary map constructed in Section~\ref{sec:binary_detection} both horizontally and vertically. Successive event regions whose lengths are close to the $1:1:3:1:1$ ratio lead to a candidate FP center at their midpoint.
We retain only candidates for which the ratio holds in both directions, which excludes false candidates within the payload and yields the four FP centers with subpixel accuracy.
The FPs are placed at fixed cell coordinates by design, and the positions of their centers on the marker are therefore known.

\subsection{Ternary Symbol Decision}\label{sec:decision}
Let $t$ be the decoding time.
We normalize $S_p(\mathbf{u};t)$ in \eqref{eq:sae} to $[0,1]$ over the time window $[t-\Delta t,\,t]$, which gives the matrix $\hat{\mathbf{S}}_p\in[0,1]^{H\times W}$: A recent event of polarity $p$ has a value close to $1$.

Using the four FP centers of Section~\ref{sec:detection} as correspondences, we project the marker region onto a rectified square region $\Omega'$.
We denote the projected signal by $\hat{\mathbf{S}}'_p$.

Let $\Omega'_{kl}\subset\Omega'$ denote a set of pixels of the cell in row $k$ and column $l$.
The number of pixels at which an event of polarity $p$ is observed in this cell is
\begin{equation}
 E_{p}(k,l)=\!\!\sum_{\mathbf{u}\in\Omega'_{kl}}\!\!\bigl[\hat{\mathbf{S}}'_p(\mathbf{u})>\theta_b\bigr],
 \label{eq:count}
\end{equation}
where $[\cdot]$ equals $1$ when the condition is true and $0$ otherwise, and $\theta_b$ is the binarization threshold.
The tentative trit value $(\hat{\mathbf{T}})_{k,l}$ of the $(k,l)$ cell is determined using a threshold $\theta_c$ as
\begin{equation}
 (\hat{\mathbf{T}})_{k,l}=
 \begin{dcases}
  +1, & \text{if } E_{+1}(k,l)\!>\!\theta_c|\Omega'_{kl}| \land E_{+1}(k,l)\!>\!E_{-1}(k,l) \\
  -1, & \text{if } E_{-1}(k,l)\!>\!\theta_c|\Omega'_{kl}| \land E_{-1}(k,l)\!\ge\!E_{+1}(k,l) \\
  0, & \text{otherwise}.
 \end{dcases}
 \label{eq:trit}
\end{equation}

\subsection{Demodulation Based on Frame-Transition Detection}\label{sec:demod}
If the decision in \eqref{eq:trit} is made asynchronously to the LCD blinking, events from several frame transitions are mixed and cause decision errors.
This is because an LCD updates its frame progressively: A frame transition proceeds from the top row of the marker to the bottom row.
The boundary between refreshed and not-yet-refreshed rows (update boundary) moves downward.

\begin{figure}[!t]
 \centering
 \includegraphics[width=\columnwidth]{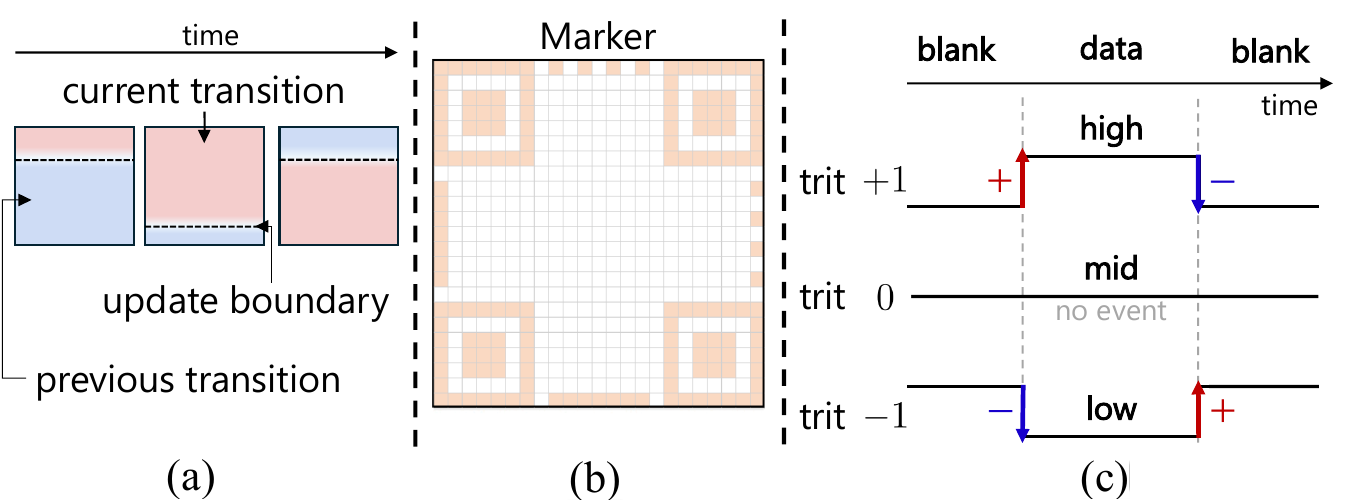} \caption{Ingredients for the proposed frame-transition detection. (a) The update boundary moves downward as a transition proceeds. (b) Orange-colored cells, whose brightness changes at every frame transition. (c) Brightness of a cell with trit $+1$, $0$, or $-1$ over the blank-to-data and data-to-blank transitions, and the polarities of the resulting events.}
 \label{fig:rowwise}
 \vspace{-6pt} \end{figure}

As a result, the events observed at a time slot may come from two successive frame transitions, as shown in Fig.~\ref{fig:rowwise}(a).
To address this, we determine the output trit values from \eqref{eq:trit} in the following two steps.
Since the temporal resolution of LCDs is significantly lower than that of EVSs, the decoding stage operates $n$ times per data-frame period $2/f$.

\subsubsection{Frame-Transition Detection}\label{sec:identify}
In the proposed marker, the orange-colored cells in those of the FPs and interior locator, shown in Fig.~\ref{fig:rowwise}(b), change their brightness at every frame transition regardless of the transmitted data\footnote{The asymmetric structure of the interior locator helps to estimate the marker rotation correctly.}\nobreak.
We use them to detect frame transitions.
Let $\mathcal{R}_k$ denote a set of the column indices of the orange-colored cells in row $k$.
To detect the frame transitions, we compute the majority event count as follows:
\begin{equation}
 g_k(t)=\frac{1}{|\mathcal{R}_k|}\sum_{l\in\mathcal{R}_k}\bigl(E_{+1}(k,l;t)-E_{-1}(k,l;t)\bigr),
 \label{eq:srow}
\end{equation}
\noindent where $E_p(k,l;t)$ is $E_p(k,l)$ in \eqref{eq:count} at the decoding time $t$.
Positive and negative values of $g_k(t)$ indicate the blank-to-data and data-to-blank transitions, respectively.
We threshold $|g_k(t)|$ with $\theta_w$ to obtain the frame-transition times of row $k$.
For each data frame, the blank-to-data transition time $\tau_0(k)$ is the first decoding time at which $g_k(t)>\theta_w$, and the data-to-blank transition time $\tau_1(k)$ is the first decoding time after $\tau_0(k)$ at which $g_k(t)<-\theta_w$.

\subsubsection{Output Trit Decision}\label{sec:combine}
We further use $\tau_0(k)$ and $\tau_1(k)$ to determine the output trit from $(\hat{\mathbf{T}})_{k,l}$.
Note that, for each data frame, we have the blank-to-data transition and the data-to-blank transition to represent one trit value, as shown in Fig.~\ref{fig:rowwise}(c).
The output trit array $\mathbf{T}$ is thus given by
\begin{equation}
 \begin{gathered}
 (\mathbf{T})_{k,l}=
 \begin{dcases}
  +1, & \text{if } \exists t',t''\colon (\hat{\mathbf{T}}(t'))_{k,l}=+1 \land (\hat{\mathbf{T}}(t''))_{k,l}=-1 \\
  -1, & \text{if } \exists t',t''\colon (\hat{\mathbf{T}}(t'))_{k,l}=-1 \land (\hat{\mathbf{T}}(t''))_{k,l}=+1 \\
  0, & \text{otherwise},
 \end{dcases}\\
 t'\in[\tau_0(k),\,\tau_1(k)),\quad t''\in[\tau_1(k),\,\tau_1(k)+\Delta t_1],
 \end{gathered}
 \label{eq:tfin}
\end{equation}
where $\hat{\mathbf{T}}(t)$ is the tentative trit array in \eqref{eq:trit} at $t$, and $\Delta t_1$ is a small constant determined by the response time of the EVS.

\begin{figure}[t]
 \centering
 \includegraphics[width=0.58\columnwidth]{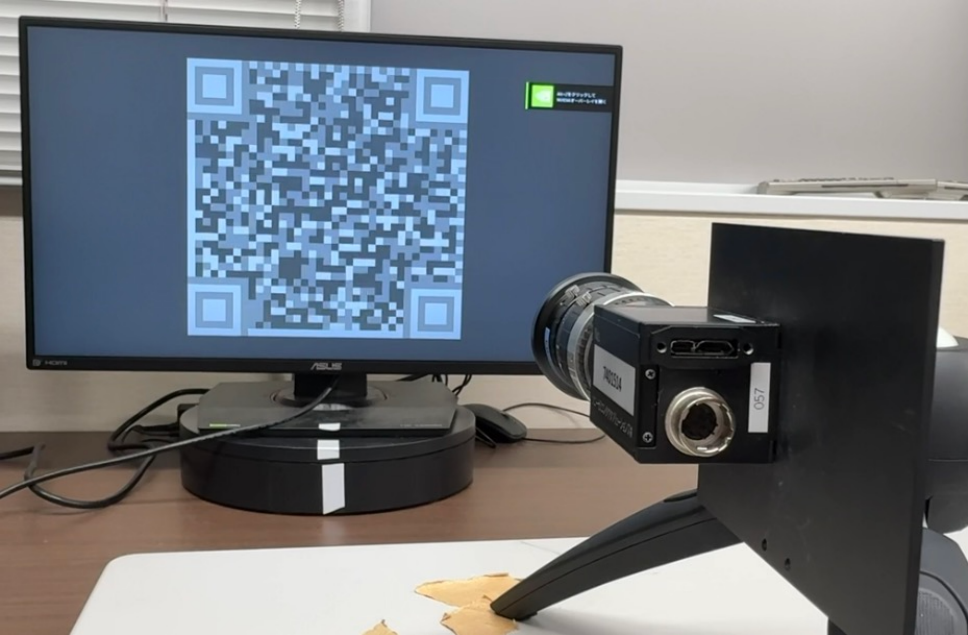} \caption{Experimental setup. The LCD on the left displays the ternary dynamic marker, and the EVS on the right captures it.}
\label{fig:env}
 \vspace{-6pt} \end{figure}

\section{Experiments}\label{sec:exp}
This section evaluates the transmission performance of the proposed method against the binary-value-based OCC using an EVS~\cite{Motion}.

\subsection{Experimental Setup}\label{sec:setup}
Fig.~\ref{fig:env} shows the experimental setup. We used an IMX636 EVS ($1280\times720$ pixels) with a FUJINON HF6XA-5M lens as the receiver and an ASUS TUF Gaming VG259QM LCD (IPS, $1920\times1080$) as the transmitter, and positioned the EVS $70~\mathrm{cm}$ in front of the display.
We rendered the marker using Psychtoolbox-3 with a monitor refresh rate of 60 Hz and a marker frame rate of 60 fps.

The asymmetric response times of LCDs to brightness increases and decreases~\cite{elze2012temporal} result in a latency difference between positive and negative event responses.
To reduce this latency difference, we set the pixel values for high, intermediate, and low brightness to $130$, $90$, and $40$, respectively.

The refractory filter uses a time window of 1 ms.
The spatiotemporal filter uses a spatial radius of 5 pixels, a time window of 10 ms, and a minimum event count of 3.
The demodulation parameters are $\theta_w=0.5$, $\theta_b=0.1$, $\theta_c=0.5$, and $n=9$.

\begin{table}[t]
 \centering
 \caption{Compared methods. FTD denotes frame-transition-based demodulation, and FP a finder pattern.}
 \label{tab:variants}
 \small
 \setlength{\tabcolsep}{4pt}
 \begin{tabular}{lccc}
  \toprule
  Method & Encoding & Localization & Demodulation \\
  \midrule
  Ternary-FTD-FP & Ternary & FP detection & FTD \\
  Ternary-FTD & Ternary & Contour approx. & FTD \\
  Ternary-OR & Ternary & Contour approx. & Logical OR \\
  Binary-OR~\cite{Motion} & Binary & Contour approx. & Logical OR \\
  \bottomrule
 \end{tabular}
 \vspace{-6pt} \end{table}

\begin{figure}[t]
 \centering
 \begin{minipage}[b]{0.49\columnwidth}\centering
  \includegraphics[width=\linewidth]{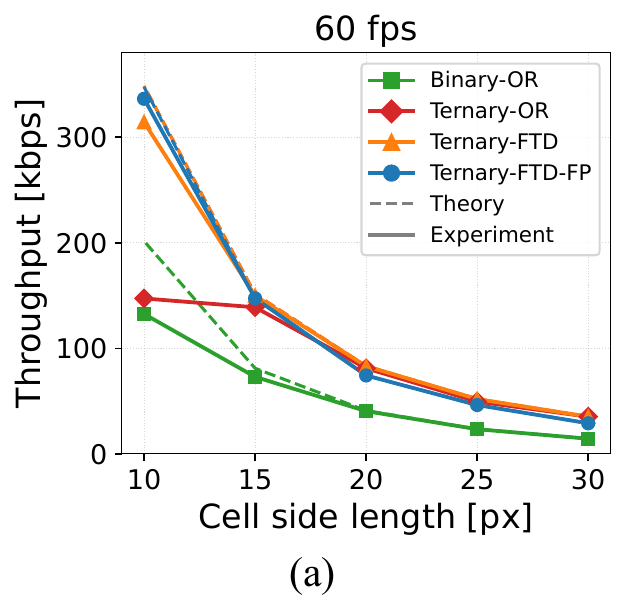}
 \end{minipage}\hfill
 \begin{minipage}[b]{0.49\columnwidth}\centering
  \includegraphics[width=\linewidth]{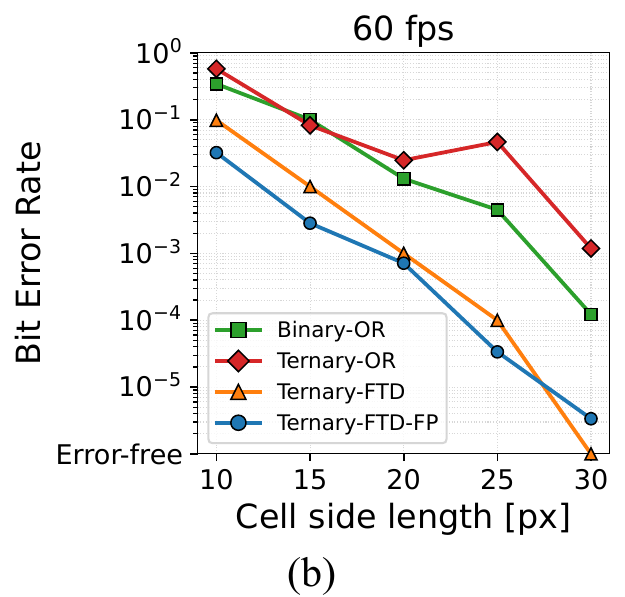}
 \end{minipage} \caption{(a) Throughput and (b) BER against the cell side length $s_c$ at 60 fps. Solid curves are averaged values over five trials. In (a), dashed curves show the maximum throughput $R_{\max}$. In (b), a zero BER is drawn on the ``Error-free'' line.}
 \label{fig:perf}
 \vspace{-6pt} \end{figure}

\subsection{Evaluation Protocol}\label{sec:protocol}\label{sec:metrics}
We use the throughput and the BER as evaluation metrics.
The throughput is $R=(1-\mathrm{BER})R_{\max}$, where $R_{\max}=\tfrac{1}{2}f\alpha N_d$ is the maximum throughput and $\alpha$ is the number of bits per payload cell.
We set $\alpha=1$ for the binary method and $\alpha=3/2$ for the ternary methods.
We compute the BER over all transmitted bits.

Table~\ref{tab:variants} lists the compared methods, including the existing binary method~\cite{Motion}, abbreviated as Binary-OR.
Ternary-FTD-FP is the full proposed method.
Ternary-FTD localizes the marker by the contour approximation of Binary-OR.
Ternary-OR also applies the same frame-transition detection as that in~\cite{Motion}, and therefore differs from Binary-OR only in the encoding.\label{sec:compared}

\label{sec:procedure}
To examine the effect of spatial resolution, we varied $s_c$ over $30$, $25$, $20$, $15$, and $10$.
The payload consists of the $N_g^2$ cells of the marker excluding those of the four FPs, the separators, and the interior locator.
The resulting number of cells along each marker side $N_g$ and the number of payload cells $N_d$ are $(N_g,N_d)=(31,645)$, $(37,1029)$, $(45,1653)$, $(61,3285)$, and $(91,7725)$.

Each trial lasted 10 s. We used the first 5 s to initialize marker detection. During the final 5 s, we converted random bit sequences to trits for the ternary methods and encoded them with the proposed marker.
We performed five trials for each condition and cell size.

\subsection{Results and Discussion}\label{sec:results}
The throughput results are shown in Fig.~\ref{fig:perf}(a).
Ternary-FTD-FP achieved the highest mean throughput of $336.5$~kbps at $s_c=10$.
Both Ternary-FTD and Ternary-FTD-FP outperformed Binary-OR for all $s_c$'s, mainly due to the change of transmitted symbols from bits to trits.

The BER results are shown in Fig.~\ref{fig:perf}(b).
In most conditions, the BER increased as $s_c$ decreased because the number of pixels per cell is decreased.
The BER of Ternary-FTD was lower than that of either OR-based method for all $s_c$'s.
This is because \eqref{eq:tfin} uses the tentative trits of the two frame transitions separately and thus avoids the errors that occur when the events observed at a time slot come from both transitions.
Ternary-FTD-FP lowered the BER further except at $s_c=30$, which suggests that detecting the four FP centers reduces the error in the estimated marker position.

We further compare reported throughputs for frame camera-based OCC systems~\cite{okawa2025high,zhang2018chromacode} in terms of bps/pixel.
The proposed method, the grid-patterned OCC system~\cite{okawa2025high}, and ChromaCode~\cite{zhang2018chromacode} have throughputs of $0.41$, $0.023$, and $0.36$~bps/pixel, respectively.
Both \cite{okawa2025high} and ChromaCode display frames at $120$~fps and use frame-based cameras.
The proposed method displays the marker at $60$~fps.
Although the experimental setups differ, these results indicate that the proposed method is comparable to the most recent frame camera-based OCCs at half the frame rate.
The BER of the proposed method, $3.2\times10^{-2}$, is also lower than that of ChromaCode, $5\times10^{-2}$.

\section{Conclusion}\label{sec:conclusion}
This paper proposed a ternary visible light communication method using an EVS and an LCD, in which each payload cell carries a ternary symbol through event polarity.
The frame-transition-based demodulation suppresses the errors caused by the transmitter--receiver asynchrony, and QR-code FPs enable subpixel-accurate localization of the reference points.
Experiments showed that the proposed method achieved a higher throughput and a lower BER than the existing binary method.

\bibliographystyle{IEEEbib}
\bibliography{refs}

\end{document}